**MAGNETODIELECTRIC EFFECT WITHOUT MULTIFERROIC COUPLING**

G. Catalan[1]

Materials Science Center, University of Groningen, Groningen 9747AG, Netherlands

&

Department of Earth Sciences, University of Cambridge, Downing Street, Cambridge CB2 3EQ, United Kingdom

**Abstract**

The existence of a magnetodielectric (magnetocapacitance) effect is often used as a test for multiferroic behavior in new material systems. However, strong magnetodielectric effects can also be achieved through a combination of magnetoresistance and the Maxwell-Wagner effect, unrelated to multiferroic coupling. The fact that this resistive magnetocapacitance does not require multiferroic materials may be advantageous for some practical applications. Conversely, it also implies that magnetocapacitance *per se* is not sufficient to establish multiferroic coupling.

[1] E-mail: gcat05@esc.cam.ac.uk

There has been a recent surge of interest in the physics and applications of multiferroics [1]. Though multiferroic materials are those where more than one ferroic order (magnetic, electric or elastic) co-exist and are coupled, the term usually refers specifically to those with ferroelectric and ferro- or antiferromagnetic order. From the applied point of view, coupling between ferroelectricity and ferromagnetism would be useful for multi-state memories, or memories with dual read-write mechanism, among other devices. From the fundamental point of view, the coexistence of ferroelectric and magnetic order also represents an interesting challenge, particularly since it has been shown that the conventional mechanism of ferroelectricity in perovskite ferroelectrics, an off-centering of B-site cations (such as $Ti^{4+}$ in $BaTiO_3$), requires the B site to have an empty $d$ orbital, which is incompatible with magnetic ordering [2].

In order to circumvent this incompatibility, two main routes are being investigated: a) materials with non-conventional mechanisms for ferroelectric and/or magnetic ordering, and b) composite materials combining conventional ferroelectrics and ferromagnetics segregated on a nanoscale level. Among the first are the so-called "geometric" multiferroics such as hexagonal $YMnO_3$ (the true nature of ferroelectricity in this compound is still subject of controversy [3-5]), highly frustrated spin systems such as $TbMnO_3$ [6] or $TbMn_2O_5$ [7], and materials combining A-site (lone pair) ferroelectricity with B-site magnetic order, such as $BiFeO_3$ [8] and $BiMnO_3$ [9]. Examples of composites are the self-segregated clusters of magnetic $CoFe_2O_4$ and ferroelectric $BaTiO_3$ [10], and superlattices combining ferromagnetic $(La,Ca)MnO_3$ with ferroelectric $BaTiO_3$ [11].

Establishing multiferroic coupling requires measuring the effect of a magnetic field on ferroelectric polarization or, conversely, that of an electric field on magnetic order. An important difficulty in doing this lies in that many candidates to be multiferroic are in fact not very good insulators, which makes it difficult for them to sustain electric fields [12]. This obstacle may be overcome by measuring depolarisation currents instead of polarisation hysteresis loops, as the former do not require the application of large electric fields. This has been done for $TbMn_2O_5$ [6] and $TbMnO_3$ [7]. Multiferroic ordering can also be detected without applying electric fields by using second-harmonic-generation [13,14], the optical frequency-doubling property of non-centrosymmetric (polar) crystals.

A relatively simple and thus widely used alternative is the examination of the dielectric constant ($\varepsilon$) as a function of temperature ($T$) or magnetic field ($H$). In a multiferroic, the dielectric constant is perturbed by the onset of magnetic ordering. Measuring $\varepsilon(T)$ and looking for deviations around the magnetic transition can therefore be used to detect multiferroic coupling, as observed in $YMnO_3$ [15] and $BiMnO_3$ [16]. Since magnetic ordering itself is affected by magnetic fields, these fields also indirectly modify the dielectric constant of multiferroics. This is the so-called magnetodielectric (or magnetocapacitance) effect. Magnetodielectric effects have been reported for several material systems, such as relaxor selenides [17], manganese oxides [18], double perovskites [19], fine-grained ferrites [20] and heteroepitaxial superlattices [11]. The potential risk with this approach, however, is that multiferroic coupling is not the only way to produce a strong magnetodielectric effect. As will be shown, magnetoresistive artefacts can also give rise to an apparently large magnetocapacitance. Thus, while multiferroicity may imply magnetocapacitance, the converse is not true.

For the (acoustic frequency) dielectric constant to be measured in a multiferroic system, a capacitor structure has to be made and an AC electric field must be applied. The response of the material to this AC field will contain at least one capacitive (dielectric) term and one resistive (leakage) term. While in good homogeneous insulators the capacitive term dominates, many would-be multiferroics are neither good insulators, nor necessarily homogeneous, and hence the measured dielectric response may be dominated or at least affected by the resistive term.

Further to this, the work-functions of electrode and dielectric material are rarely identical, so band-bending may occur near the electrode-dielectric interfaces. This induces charge injection from the electrode into the dielectric or vice-versa (charge-depletion). In either case, the result is a layer near the electrode-dielectric interface with a different density of charge carriers, and hence different conductivity. When the dielectric is not a very good insulator this may cause the electric field to be mostly dropped in the charge-depleted interfacial area rather than in the bulk of the material, producing artificially high dielectric constants. This effect has been documented in several oxide materials, including manganites [21]. By the same token, superlattices of materials with different resistivities can also have artificially high dielectric constants [22].

Whether the heterogeneous nature of the sample is deliberate (as in a superlattice), or accidental (interfacial or grain-boundary layers), either case can be described by the Maxwell-Wagner (M-W) capacitor model. This effectively consists of two leaky capacitors in series (Figure 1). The impedance of such a system under an AC field is a complex quantity, and the real and imaginary parts of its permittivity are [22, 23]:

$$\varepsilon^{'}(\omega) = \frac{1}{C_0(R_i + R_b)} \frac{\tau_i + \tau_b - \tau + \omega^2 \tau_i \tau_b \tau}{1 + \omega^2 \tau^2}$$

and

$$\varepsilon^{''}(\omega) = \frac{1}{\omega C_0(R_i + R_b)} \frac{1 - \omega^2 \tau_i \tau_b + \omega^2 \tau(\tau_i + \tau_b)}{1 + \omega^2 \tau^2},$$

where subindex $i$ and $b$ refer to the interfacial-like and bulk-like layers respectively, $R$=resistance, $C$=capacitance, $\omega$= AC frequency, $\tau_i = C_i R_i$, $\tau_b = C_b R_b$, $\tau = \dfrac{\tau_i R_b + \tau_b R_i}{R_i + R_b}$, $C_0 = \varepsilon_0 \dfrac{A}{t}$, $A$ = area of the capacitor, $t$ = thickness.

If the resistance of any of the layers is changed by a magnetic field, so will the measured capacitance. Magnetoresistance combined with the Maxwell-Wagner effect thus provides a mechanism for magnetocapacitance in materials that are not necessarily multiferroic.

It is illustrative to consider two archetypal cases. First, that of the magnetocapacitance of a simple magnetoresistive material. By way of example, let us assume that the material is a semiconductor at the core with interfacial regions whose resistivities are either 10 times higher (charge-depleted) or 10 times lower (charge-injected). Let us assume also that the interfacial regions represent 10% of the capacitor's thickness, and that there is no substantial magnetoresistance in them. The intrinsic dielectric constant, on the other hand, should be the same for both core and interface. In order to put numbers to the equations, values for the dielectric constant of 25 (typical of multiferroic manganites) and a core resistivity of $10^4 \Omega$m have been assumed, together with a negative magnetoresistance which depends on the magnetic field as $(H/H_S)^{1/2}$, being –50% at $H_S$=10T. All these parameters are plausible but are solely intended as an illustration: qualitative consequences do not substantially depend on them. Using the M-W equations it is now possible to calculate the real part of the dielectric constant, and thus also the magnetocapacitance, defined as $MC = \dfrac{\varepsilon'(H) - \varepsilon'(0)}{\varepsilon'(0)} \times 100$. The result for $\omega$=1kHz is shown in Figure 2.

Three features are particularly noteworthy. First, giant magnetoresistance can yield giant magnetocapacitance. Second, for negative magnetoresistance the sign of the magnetocapacitance can be positive or negative depending on whether the interfacial regions have bigger or smaller resistivity than the core. Finally, the shape of the magnetocapacitance as a function of magnetic field is directly related to that of the magnetoresistance, assumed here to vary as a square root of the field.

A second representative case is that of a superlattice combining a purely ferroelectric material with a purely magnetorresistive one. Here the relative dielectric constants of the two components are different, typically $\varepsilon_r \sim 250$ for barium titanate films and $\varepsilon_r \sim 25$ or less for manganites. The resistivity of ferroelectric thin films depends considerably on processing conditions, but values in the range $10^4$-$10^6$ $\Omega$m are reasonable. As for the magnetoresistive layer, resistivity depends critically on specific material, dopant density and strain. For the sake of simplicity, the magnetoresistive layers are assumed here to have the same resistivity at zero field as the ferroelectric ones ($10^5 \Omega$m). It is worth mentioning that although the magnetoresistance has been assumed to reside exclusively in the magnetoresistive layer, it is also possible to tune magnetically the size of the depletion layers at the junctions between manganite and titanate [24]. As in the previous example, these assumptions are not critical and are intended for illustrative purposes only. In Figure 3 the calculated $\varepsilon'(\omega)$ is plotted with and without applied magnetic field.

As shown in Figure 3, frequency plays an important role: at high frequencies charge carriers do not have time to respond, and the measured capacitance is simply that of two capacitors in series. At low frequencies, on the other hand, the charge carriers in the low resistivity layer have time to respond, so most of the field is dropped across the layers with bigger resistivity, resulting in an increased apparent capacitance. This frequency dependence provides a useful test: unless very slow dynamics (such as in glasses or slow domain walls) are involved, intrinsic (non conductivity related) magnetocapacitance should generally be measurable at frequencies higher than the conductivity cutoff.

These two calculations show that large ("colossal") magnetocapacitive effects can be achieved in material systems without multiferroic coupling. The model presented here requires only that within such material systems there exist magnetoresistive regions. As well as from boundary effects, heteroresistive behaviour may also arise from doping: mixed valence manganites, for example, have clusters of metallic (double-exchange) islands in a semiconducting matrix. These considerations may also apply to relaxor selenides, given their nano-structured nature and large dielectric losses [17].

To summarize, we have shown that large magnetodielectric effects can be obtained in systems combining regions of different conductivity where at least one of the component regions is magnetoresistive. This effect has the advantage that the otherwise rare multiferroic materials are not needed to achieve it (though high losses would normally be concomitant). Conversely, measuring a magnetodielectric effect is in principle insufficient to establish conclusively the existence of true multiferroic coupling, unless accompanied by careful examination of frequency dependence, dielectric loss and, where possible, magnetoresistance.

This work has been partly funded by the EU under the Marie Curie Intra-European-Fellowship programme. The author wishes to thank U. Adem, M. Gich, B. Noheda and J.F. Scott for their useful comments.

**FIGURE CAPTIONS**

**Figure 1/3 (color online)**: Capacitor systems with magnetically tunable Maxwell-Wagner behaviour: (a) homogeneous material with charge-depleted interfacial layers; (b) superlattice (c) clustered material or fine-grained ceramic with grain boundaries having different resistivity than the grain interior. All three systems are well described by two leaky capacitors in series (the Maxwell-Wagner model). If one of the components is magnetoresistive, one of the resistivities will be tunable.

**Figure 2/3**: Magnetocapacitance of a magnetorresistive material with non-magnetorresistive boundary layers

**Figure 3/3:** Real part of the dielelectric constant as a function of frequency for a superlattice consisting of a ferroelectric material and a material with negative magnetorresistance.

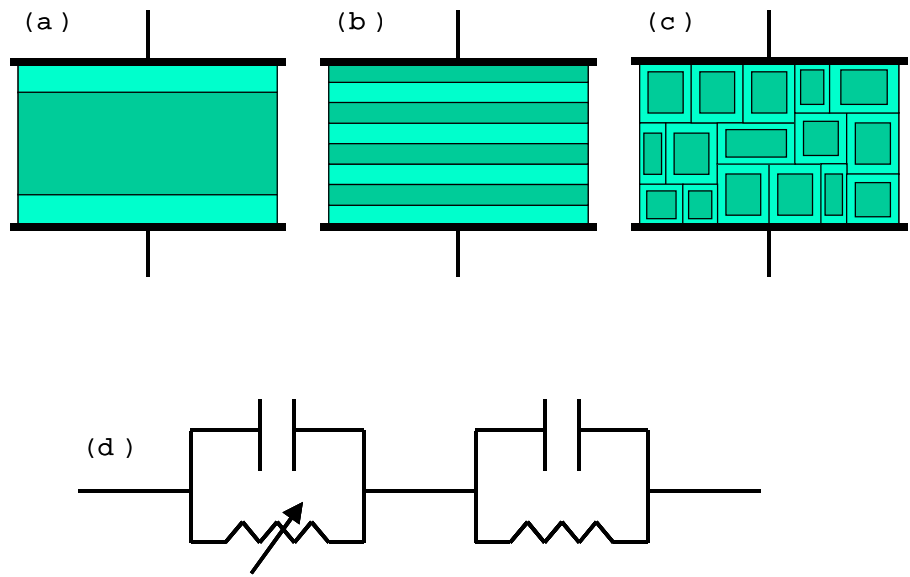

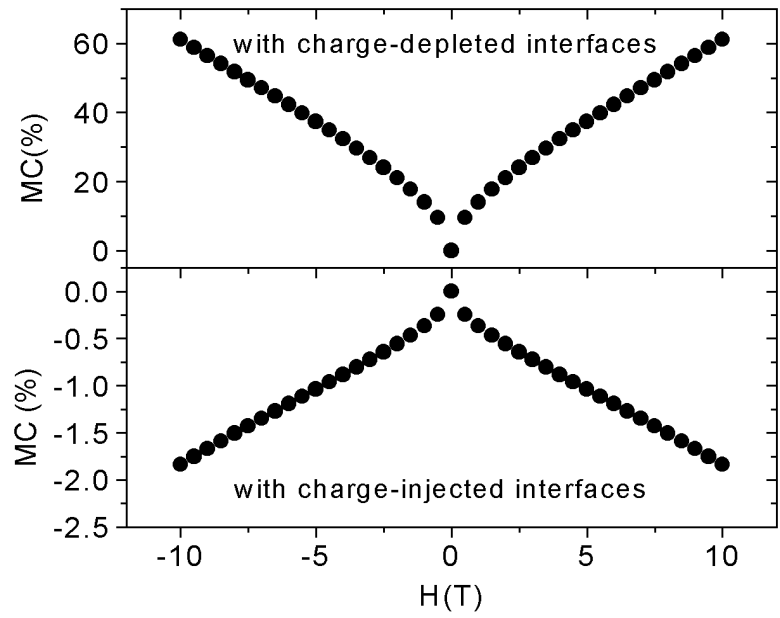

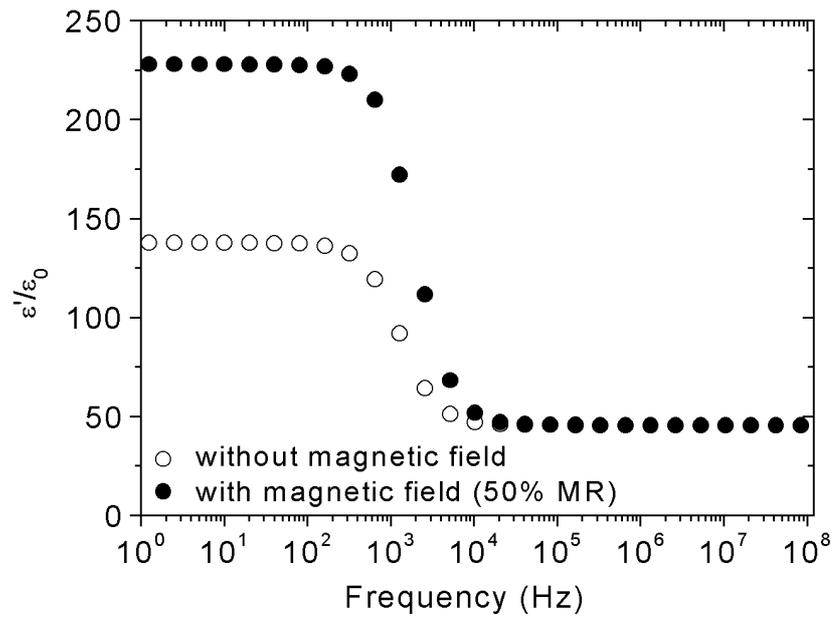